\documentclass[notitlepage, prl, aps, floatfix, twocolumn]{revtex4-1} 
\usepackage{CJK}
\usepackage{pdfsync}
\usepackage{graphicx,bm}
\usepackage{subfigure} 
\usepackage{amssymb}
\usepackage{psfrag}
\usepackage{amsmath}
\usepackage{verbatim}
\usepackage{my_symbols}
\usepackage{amsfonts}
\usepackage{graphicx}

\begin{document}
\begin{CJK*}{UTF8}{gbsn} % Use default fonts from CJK (see below)

\title{Spin wave excitation in magnetic insulators by spin-transfer torque}

\author{Jiang Xiao (萧江)}
\affiliation{Department of Physics and State Key Laboratory of Surface Physics, Fudan
University, Shanghai 200433, China}
\author{Gerrit E. W Bauer}
\affiliation{Institute for Materials Research, Tohoku University, Sendai 980-8557, Japan \\
Kavli Institute of NanoScience, Delft University of Technology, 2628 CJ Delft, The Netherlands}

\date{\today}
%=========================================================================
\begin{abstract}

	We study the excitation of spin waves in magnetic insulators by the current-induced
	spin-transfer torque. We predict preferential excitation of surface spin waves induced by
	an easy-axis surface anisotropy with critical current inversely proportional to the
	penetration depth and surface anisotropy. The surface modes strongly reduce the critical
	current and enhance the excitation power of the current-induced magnetization dynamics.

\end{abstract}
%=========================================================================

\maketitle
\end{CJK*}

%###############################################################

%>>>>>>>>>>>>>>>>>>>>>>>>>>>>>>>>>>>>>>>>>>>>>>>>>>>>>>>>>>>>>>>>>>>>>>>>>

%{\it Introduction:} 
Spintronics is all about manipulation and transport of the spin, the intrinsic angular momentum
of the electron \cite{zutic_spintronics:_2004}. These two tasks are incompatible, since
manipulation requires strong coupling of the spin with the outside world, which perturbs
transport over long distances. In normal metals spin can be injected and read out easily, but
the spin information is lost over short distances \cite{bass_spin-diffusion_2007}.
%The world record for the spin flip length is in Ag [kimura] and in graphene [van Wees]. 
In spin-based interconnects, transporting spins over longer distances is highly desirable
\cite{khitun_non-volatile_2011}.

The long-range transport of spin information can be achieved by encoding the information into
spin waves that are known to propagate coherently over centimeters \cite{serga_yig_2010}. It
has been demonstrated in Refs. \onlinecite{kajiwara_transmission_2010, madami_direct_2011,
wang_control_2011} for the magnetic insulator Yttrium-Iron-Garnet (YIG) that spin waves can be
actuated electrically by the spin-transfer torque \cite{slonczewski_current-driven_1996,
berger_emission_1996} and detected by spin pumping \cite{tserkovnyak_enhanced_2002} at a
distant contact. In the experiment by Kajiwara \cite{kajiwara_transmission_2010}, Pt was used
as spin current injector and detector, making use of the (inverse) spin Hall effect
\cite{saitoh_conversion_2006}. In a $d = 1.3~{\mu}$m-thick YIG film spin waves were excited by a
threshold charge current of $J_c {\sim} 10^9$ A/m$^2$. This value is much less than expected for
the bulk excitation that in a linear approximation corresponds to the macrospin mode and is
estimated as $J_c = (1/{\theta}_H) e{\alpha}{\omega}M_sd/ {\gamma}{\hbar}{\sim} 10^{11{\sim}12}$ A/m$^2$, where $e$ and ${\gamma} $ are
the electron charge and gyromagnetic ratio, respectively, and we used the parameter values in
Table \ref{tab:param} for the ferromagnetic resonance frequency ${\omega}$, the spin Hall angle of Pt
${\theta}_H$, magnetic damping ${\alpha}$, and saturation magnetization $M_s$.
%In this estimate we disregarded any reflection of the spin current based on recent
%insights on the NM$|$YIG interface [Jia, Heinrich APL, submitted]. 

In this Letter, we address this large mismatch between observed and expected critical currents
by studying the threshold current and excitation power of current-induced spin wave
excitations. We present a possible answer to the conundrum by proving that the threshold
current is strongly decreased in the presence of an easy-axis surface anisotropy (EASA).
Simultaneously, EASA increases the power of the spin wave excitation by at least two orders of
magnitude.

\begin{figure}[t]
	\includegraphics[width=0.4\textwidth]{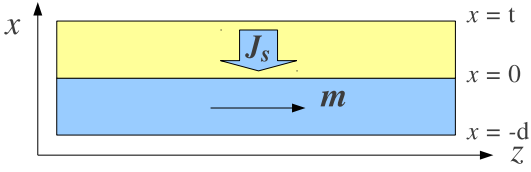}
	\caption{(Color online) An electrically insulating magnetic film of thickness $d$ with
	magnetization $\mm$ ($\|\hzz$ at equilibrium) in contact with a normal metal. A spin
	current $J_s\|\hzz$ is generated in the normal metal and absorbed by the ferromagnet.}
	\label{fig:NF}
	\vspace{-0.8cm}
\end{figure}

%=========================================================================
%{\it Method}: 
We study a structure as depicted in \Figure{fig:NF}, where a non-magnetic (N) metallic thin
film of thickness $t$ is in contact with a ferromagnetic insulator (FI), whose equilibrium
magnetization is along the $z$-direction. The spin current injected into the ferromagnetic
insulator is polarized transverse to the magnetization $\JJ_s = J_s \mm {\times} \hzz {\times} \mm$.
%The spin accumulation at $x = 0$ induced by a uniform spin current polarized along the
%$\hzz$-direction $\JJ_s = J_s\hzz$ flowing in $-\hxx$ direction is approximately ${\delta}{\bm {\mu}} {\simeq}
%(2e^2{\rho}_N{\lambda}_N/{\hbar})J_s\hzz {\simeq} (3\sqrt{3}{\pi}^2/k_F^2\bar{{\eta}}_{\rm so})J_s\hzz$, where ${\rho}_N, {\lambda}_n, k_F,
%\bar{{\eta}}_{\rm so}$ are the electrical resistivity, spin-diffusion length, Fermi wave-vector, and
%the dimensionless spin-orbit coupling parameter of N, respectively, as defined in Ref.
%\onlinecite{takahashi_spin_2008}. The spin-transfer torque at the N/FI interface due to
%${\delta}{\bm{\mu}}$ is ${\bm {\tau}} = (g_r/4{\pi}A)\mm{\times}{\delta}{\bm {\mu}}{\times}\mm$ with $g_r/A {\sim} k_F^2/4{\pi}$ the mixing
%conductance per area for perfect interface. Therefore ${\bm{\tau}}{\simeq} {\xi} J_s\mm{\times}\hzz{\times}\mm$ with ${\xi} =
%3\sqrt{3}/16\bar{{\eta}}_{\rm so} {\sim} 1$.
%Its transverse component is absorbed at the
%interface thus exerting a torque on the magnetization $\mm$ \cite{stiles_anatomy_2002}. 
The bulk magnetization is described by the Landau-Lifshitz-Gilbert (LLG) equation:
%-----------------------------------------
\begin{equation}
%\label{eqn:llgh}
%\begin{align}
	\label{eqn:llg}
%	\dmm = -{\gamma}\mm{\times}\smlb{\HH_0 + {A_{\rm ex}\ov {\gamma}}{\nabla}^2\mm + \hh} + {\alpha}\mm{\times}\dmm, 
	\dmm = -{\gamma}\mm{\times}\midb{\HH_0 + (A_{\rm ex}/ {\gamma}){\nabla}^2\mm + \hh} + {\alpha}\mm{\times}\dmm, 
%	0 &= {\nabla}{\times}\hh, \\
%	0 &= {\nabla}{\cdot}\bb = {\nabla}{\cdot}(\hh+{\mu}_0M_s\mm),
%\end{align}
\end{equation}
%-----------------------------------------
where $\HH_0$ includes the external and internal magnetic field, $A_{\rm ex}$ is the exchange
constant, and $\hh$ is the dipolar field that satisfies Maxwell's equations. In the quasistatic
approximation, {\it i.e.} disregarding retardation in the electromagnetic waves, ${\nabla}{\times}\hh = 0$
and ${\nabla}{\cdot}\bb = {\nabla}{\cdot}(\hh+{\mu}_0M_s\mm) = 0$. All quantities are position and time dependent.
In the absence of pinning, the total torque vanishes at the interface
\cite{gurevich_magnetization_1996}:
%-----------------------------------------
\begin{equation}
	A_{\rm ex}\mm{\times}{{\partial}\mm\ov{\partial}\nb} - {2{\gamma}K_s\ov M_s}(\mm{\cdot}\nb)\mm{\times}\nb + {{\gamma}J_s\ov M_s}\mm{\times}\hzz{\times}\mm = 0, 
%	\mm{\times}{{\partial}\mm\ov{\partial}\nb} - k_s(\mm{\cdot}\nb)\mm{\times}\nb + k_j\mm{\times}\hzz{\times}\mm = 0, 
\label{eqn:bc}
\end{equation}
%-----------------------------------------
where $\nb$ is the outward normal as seen from the ferromagnet. The first term in \Eq{eqn:bc}
is the surface exchange torque, the second term the torque due to a perpendicular uniaxial
surface anisotropy $\HH_a = {2K_1\ov M_s}(\mm{\cdot}\nb)\nb$ and $K_s = {\int}dx K_1$ across the
surface, and the last term is the current-induced spin-transfer torque
\cite{stiles_anatomy_2002}. We parameterize the surface anisotropy and spin current as wave
numbers $k_s = 2{\gamma}K_s/A_{\rm ex}M_s$ and $k_j = {\gamma}J_s/A_{\rm ex}M_s$. The dipolar fields
$h_{y,z}$ and $b_x$ are continuous across the interface. \Eqs{eqn:llg}{eqn:bc} in combination
with Maxwell's equations describe the low energy magnetization dynamics and can be transformed
into a 6th-order differential equation for the scalar potential ${\psi}$ with $\hh = - {\nabla}{\psi}$
\cite{de_wames_dipole-exchange_1970, hillebrands_spin-wave_1990}.

%-----------------------------------------
%-----------------------------------------
%-----------------------------------------
\begin{table}[b]
%	\centering
	\begin{tabular}{l|l|l} \hline
		Parameter				& YIG				& Unit 		\\ \hline  
		${\gamma}$ 					& $1.76{\times}10^{11}$ 	& 1/(T s) 	\\        
		$M_s$					& $^a1.56{\times}10^5$		& A/m 		\\          
		${\omega}_M={\gamma}{\mu}_0M_s$			& $34.5$			& GHz 		\\          
%		$D_{ex}$				& $5.0{\times}10^{-40}$	& J m$^2$ 	\\      
		$A_{ex}$				& $4.74{\times}10^{-6}$	& m$^2$/s 	\\      
		${\alpha}$						& $^a6.7{\times}10^{-5}$	& - 		\\              
		${\omega}_0 = {\gamma}H_0$			& $0.5{\omega}_M$			& GHz 		\\          
		$K_s$					& $^b5{\times}10^{-5}$  	& J/m$^2$	\\ \hline
%		$K_s$					& $^b-5.2{\times}10^{-5}$  & J/m$^2$	\\ \hline
%		$k_s$					& $-2.5{\times}10^7$		& 1/m 		\\          
%		${{\gamma}{\hbar}\ov 2eA_{ex}M_s}$	& $7.8{\times}10^{-5}$		& m/A 		\\ \hline   
%		${\theta}_H$					& -					& - 		\\ \hline
	\end{tabular}
	\caption{Parameters for YIG.
		$^a$Ref. \onlinecite{kajiwara_transmission_2010},
		$^bK_s$ ranges $0.01 {\sim} 0.1$ erg/cm$^2$ or $10^{-5} {\sim} 10^{-4}$ J/m$^2$, Ref.
		\onlinecite{yen_magnetic-surface_1979, ramer_effects_1976}.
%		$^b$Ref. \onlinecite{borghese_damped_1980},
		}
	\label{tab:param}
\end{table}
%-----------------------------------------

The method described above extends a previous study by Hillebrands
\cite{hillebrands_spin-wave_1990} by including the current-induced spin-transfer torque. We
predict the critical conditions under which magnetization dynamics becomes amplified by the
current-induced driving torque.

%=========================================================================
%{\it Results:} 
We start with the limiting case of $d\ra {\infty}$ (semi-infinite ferromagnet). After linearization
and Fourier transformation in both time and space domains, \Eq{eqn:llg} reduces to
%%-----------------------------------------
%\begin{equation*}
%	\smatrix{\hat{{\Omega}} & i{\omega} \\ -i{\omega} & \hat{{\Omega}}}\smatrix{m_x({\omega}) \\ m_y({\omega})} = 0
%	\qwith \hat{{\Omega}} = {\omega}_0 + i{\alpha}{\omega} - A_{\rm ex}{\nabla}^2,
%\end{equation*}
%%-----------------------------------------
%which is 
a 4th-order differential equation in ${\psi}$. Focusing for simplicity first on the case of
vanishing in-plane wave-vector $\qq = (q^y, q^z) = 0$, the scalar potential can be written as:
${\psi}(\rr) = {\sum}_{j=1}^2a_je^{iq_jx}e^{i{\omega}t}$ with
%-----------------------------------------
\begin{equation}
q_j({\omega}) = -i\smlb{{\omega}_0+\half {\omega}_M {\pm} \sqrt{{\omega}^2+\quarter{\omega}_M^2} + i{\alpha}{\omega}\over A_{\rm ex}}^{\half}
\end{equation}
%-----------------------------------------
and $\abs{q_1}{\gg} \abs{q_2}$ when ${\omega} {\sim} {\omega}_0$. Imposing the boundary condition in
\Eq{eqn:bc}, up to the first order in $k_j$:
%%-----------------------------------------
%\begin{equation*}
%	\left. \smatrix{{\partial}_x + k_s & -k_j \\ k_j & {\partial}_x } 
%	\smatrix{ -ie^{iq_1x} & ie^{iq_2x} \\
%	  e^{iq_1x} &  e^{iq_2x} }
%	\smatrix{a_1 \\ a_2} \right|_{x = 0}
%	= 0, 
%\end{equation*}
%%-----------------------------------------
%-----------------------------------------
\begin{equation}
%	2iq_1q_2 -(q_1+q_2)k_s + 2i(q_1-q_2)k_j = 0
	0 = 2q_1q_2(q_1+q_2) 
	+ik_s\midb{(q_1+q_2)^2+{{\omega}_M\ov A_{\rm ex}}} 
	+ 4k_j{\omega}.
%	q_j &= -i\midb{({\omega}_0+{\omega}_M/2{\pm}\sqrt{{\omega}^2+{{\omega}_M^2\ov 4}}+i{\alpha}{\omega})/A_{\rm ex}}^\half.
\label{eqn:det}
\end{equation}
%-----------------------------------------
The solutions of \Eq{eqn:det} are the {\it complex} eigen-frequencies ${\omega}$, whose real part
represents the energy and imaginary part the inverse lifetime. To 0th-order in dissipation,
{\it i.e.} with vanishing bulk damping (${\alpha} = 0$) and spin current injection ($k_j = 0$), and
using $\abs{q_1} {\gg} \abs{q_2}$, \Eq{eqn:det} simplifies to $k_s = iq_2/[1+{\omega}_M/(Aq_1^2)]$,
which has no non-trivial solution for $k_s \le 0$. The single real solution for $k_s > 0$ obeys
${\omega} < \sqrt{{\omega}_0({\omega}_0+{\omega}_M)}$ such that both $q_{1,2}$ are negative imaginary: $q_1 {\simeq}
-i\sqrt{(2{\omega}_0+{\omega}_M)/A_{\rm ex}}, q_2 {\simeq} -ik_s{\omega}_0/(2{\omega}_0+{\omega}_M) + O(k_s^2)$, {\it i.e.} a
surface spin wave induced by the easy-axis surface anisotropy. With the criteria Im$~{\omega} < 0$
and to leading order in $0 < k_s{\ll} q_1$, \Eq{eqn:det} leads to the critical current:
%-----------------------------------------
\begin{equation}
	k_j^c {\approx} -{{\alpha}\ov k_s}{ ({\omega}_0+{\omega}_M/2)^2\ov A_{\rm ex}{\omega}_0} 
	+ {\alpha} {{\omega}_0+2{\omega}_M\ov 4{\omega}_0}\sqrt{2{\omega}_0+{\omega}_M\ov A_{\rm ex}}.
\label{eqn:kjc}
\end{equation}
%-----------------------------------------
When there is no surface anisotropy ($k_s \ra 0$), the critical current diverges because the
macrospin mode cannot be excited in a semi-infinite film. Using the parameters given in Table
\ref{tab:param} in \Eq{eqn:kjc}, we estimate the critical current for exciting the EASA induced
surface wave (at $\qq = 0$) to be $k_j^c = -0.08k_c$, where $k_c = {\alpha}({\omega}_0+{\omega}_M/2)d/A_{\rm ex}$
is the critical current for bulk excitation in a YIG thin film of thickness $d = 0.61~{\mu}$m
(used below).
%(equivalent to $J_c {\sim} (1/{\theta}_H)3.8{\times}10^9$A/m$^2$)

EASA pulls down a surface spin wave for the following reason: when $k_j = J_s = 0$, the
boundary condition in \Eq{eqn:bc} requires cancellation between the exchange and surface
anisotropy torques: ${\partial}_xm_x - k_sm_x = {\partial}_xm_y = 0$. The exchange torque depends on the
magnetization derivative in the normal direction, and can only take one sign in the whole film,
and $m_{x,y} \ra 0$ as $x\ra -{\infty}$, therefore $(1/m_x){\partial}_x m_x > 0$. Torque cancellation (for
a non-trivial solution) is therefore possible only for $k_s > 0$. The surface spin wave induced
by EASA ($k_s > 0$) for the {\it in-plane} magnetized film ($m_z {\sim} 1$) discussed in this
Letter is analogous to the surface spin waves for the {\it perpendicular} magnetized film ($m_x
{\sim} 1$) induced by easy-plane surface anisotropy ($k_s < 0$) studied before in YIG films \cite{
%puszkarski_theoretically_1972, 
puszkarski_surface_1973,
%yu_tensorial_1974,
%yu_exchange-dominated_1975, 
wigen_microwave_1984, patton_magnetic_1984, kalinikos_theory_1986,
gurevich_magnetization_1996}. For perpendicular magnetization, a different boundary condition:
${\partial}_xm_{y,z}+k_sm_{y,z} = 0$ results in a surface wave for $k_s < 0$.

We now include all ingredients: finite thickness ($d = 0.61~{\mu}$m), surface anisotropy,
intrinsic magnetic damping, spin current injection, exchange coupling, and dipolar fields. We
calculate numerically the complex eigen-frequencies ${\omega}(\qq, k_j)$ as a function of the
in-plane wave-vector $\qq$ and the applied spin current at the surface $k_j$. Im$~{\omega}$, the
effective dissipation, can be either positive (damping) or negative (amplification) when driven
by the spin-transfer torque.

%-----------------------------------------
\begin{figure}[t]
	\includegraphics[width=0.48\textwidth]{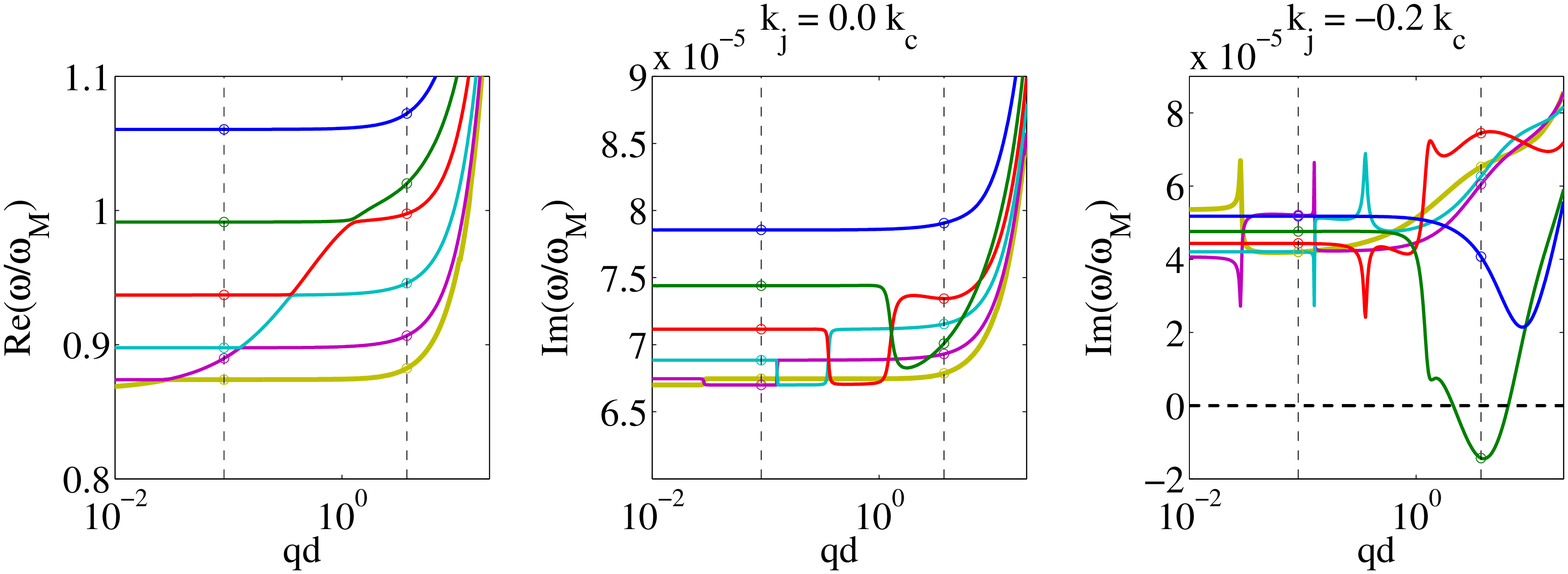}
	\includegraphics[width=0.5\textwidth]{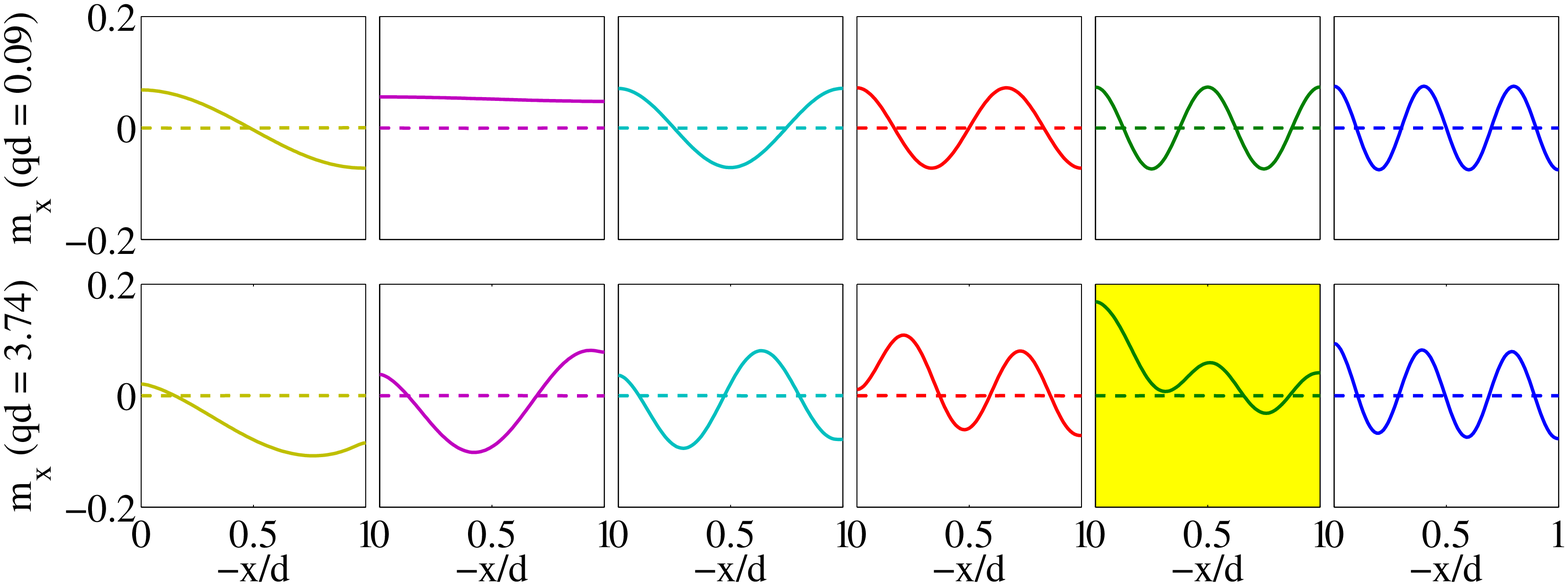}
	\caption{(Color online) Spin wave band structure and magnetization profiles in YIG for $d =
	0.61~{\mu}$m without surface anisotropy: $k_s = 0$ at ${\theta} = {\angle}(\mm,\qq) = 90^o$. Top (from left
	to right): $\re{{\omega}/{\omega}_M}$ vs.  $qd$, $\im{{\omega}/{\omega}_M}$ at $k_j = 0$, $\im{{\omega}/{\omega}_M}$ at $k_j =
	-0.2k_c$. Bottom: $m_x$ of the same 6 modes for $qd = 0.09$ and $3.74$ indicated by the
	dashed vertical lines in the top panels. The colors label different bands. The amplitude of
	the green mode mode (shaded/yellow panel) is amplified.}
%	\includegraphics[height=2.7in]{fig/ks0l}
%	\hspace{0.2in}
%	\includegraphics[height=2.7in]{fig/ks0c}
%	\caption{Left: Dispersion relation ${\omega}$ vs. $\qq d$ for the first 5 bands with $d = 0.61{\mu}$m,
%	$k_s = 0$ (no surface anisotropy), ${\theta} = {\angle}(\mm,\qq) = -60^o$. Middle and Right: the full
%	angular dependence for the 2nd and 3rd band. The 1st row: $\re{{\omega}/{\omega}_M}$, the 2nd row:
%	$\im{{\omega}/{\omega}_M}$ for $k_j = 0$, the 3rd row: $\im{{\omega}/{\omega}_M}$ for $k_j = 0.2k_c$. } 
	\label{fig:ks0}
%	\vspace{-0.5cm}
\end{figure}
%-----------------------------------------

%Dispersion without surface anisotropy.
First, we disregard the surface anisotropy: $K_s = k_s = 0$. With ${\theta}$ the angle between $\qq$
and $\mm$, the results for ${\theta} = 90^o$ are shown in \Figure{fig:ks0}. In the top left panels
Re$~{\omega}$, the magnetostatic surface wave (MSW) is seen to cross the flat bulk bands
\cite{de_wames_dipole-exchange_1970}. When no spin current is applied ($k_j = 0$), the
dissipative part Im$~{\omega} {\sim} {\alpha}({\omega}_0+{\omega}_M/2) > 0$, as shown in the top middle panels. At a spin
current that is 20\% of that required for bulk excitation: $k_j = 0.2k_c$, the dissipative part
Im$~{\omega}$ (top right panel) decreases while Re$~{\omega}$ remains unchanged because the spin-transfer
torque as magnetic (anti-)damping mainly affects Im$~{\omega}$. Negative effective dissipation
implies spin wave amplification. This happens for the 5th (green) band at $qd {\in} [2,6.5]$,
which corresponds to a (chiral) MSW (mixed with bulk modes) formed near the interface
(shaded/yellow panel). On the other hand, for ${\theta} = -90^o$ (not shown), the magnetostatic
surface wave at the opposite surface to vacuum ($x = -d$) is only weakly affected by the spin
current injection at $x = 0$.

\begin{figure}[t]
	\includegraphics[width=0.48\textwidth]{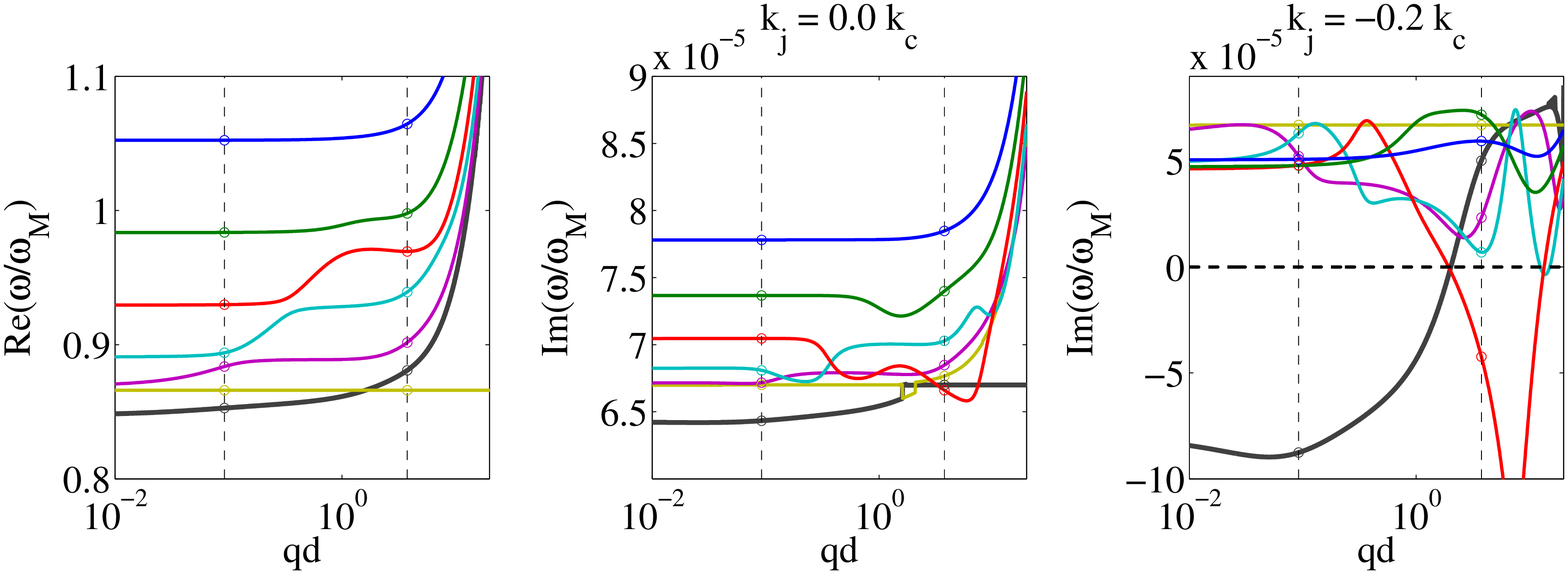}
%	\hspace{.2in}
	\includegraphics[width=0.5\textwidth]{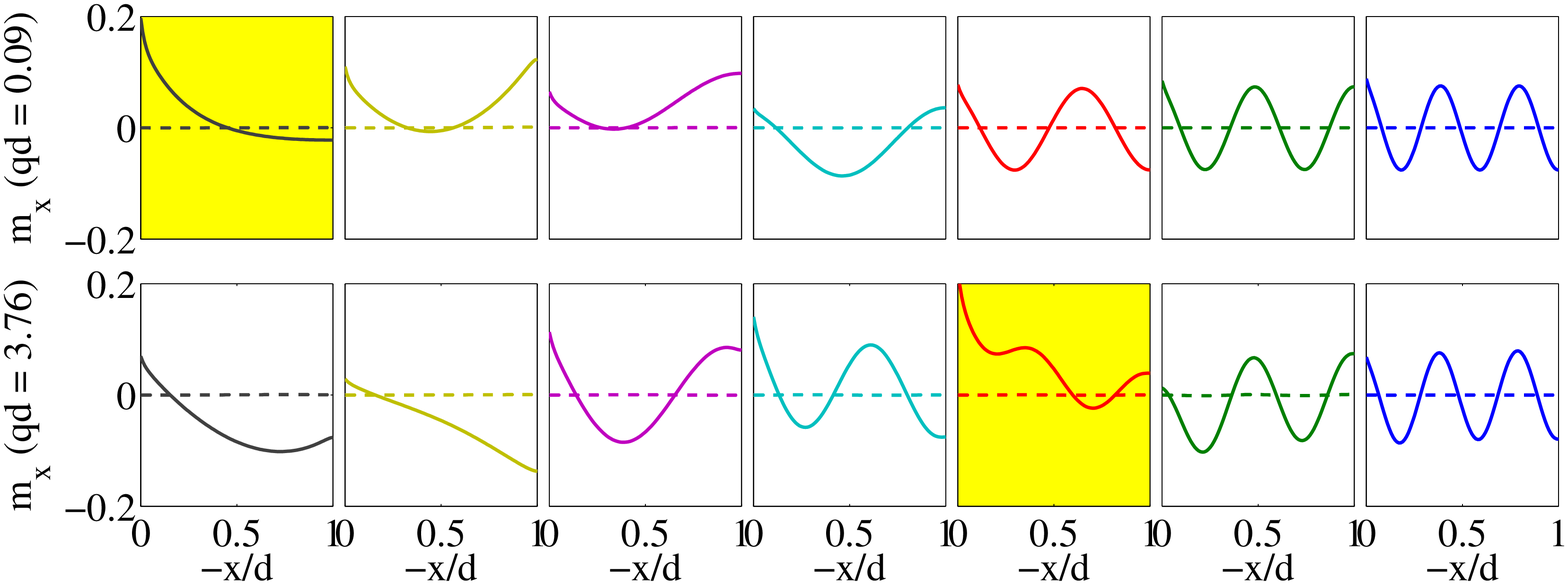}
	\caption{(Color online) Same as \Figure{fig:ks0} but with $k_s = 25/{\mu}$m.} 
	\label{fig:ks1}
%	\vspace{-0.5cm}
\end{figure}

%Dispersion with surface anisotropy.
We now turn on EASA: $k_s = 25.0/{\mu}$m (or $K_s = 5{\times}10^{-5}$J/m$^2$) at the top surface ($x =
0$). \Figure{fig:ks1} shows the results for ${\theta} = 90^o$. The changes of Re$~{\omega}$ and Im$~{\omega}$ at
$k_j = 0$ are modest (\Figure{fig:ks0}), but an additional band (black) appears, {\it viz}. the
surface spin wave band induced by EASA. The spin-transfer torque strongly affects this mode
because of its strong surface localization \cite{sandweg_enhancement_2010}. As seen in the top
right panel, almost the whole band is strongly amplified by a spin current injection of $k_j =
0.2k_c$. Inspecting the spin wave profiles at two different $q$ values, we observe a surface
spin wave near $x = 0$ for the black band at small $q$ (shaded/yellow panel in the middle row
in \Figure{fig:ks1}).
%The surface wave anti-crosses with several magnetostatic waves. 
At larger $q$, the 1st (black) band loses its surface wave features to the 5th (red) band (see
top right panel in \Figure{fig:ks1}). The red band mode starts out as a magnetostatic surface
spin wave, but the EASA enhances its surface localization by hybridization with the black mode
to become strongly amplified by the spin current at higher $q$. Also in the lower panel of
\Figure{fig:ks1} we observe that the red band has acquired the surface character.

%-----------------------------------------
%-----------------------------------------

%-----------------------------------------
\begin{figure}[b]
	\includegraphics[width=0.5\textwidth]{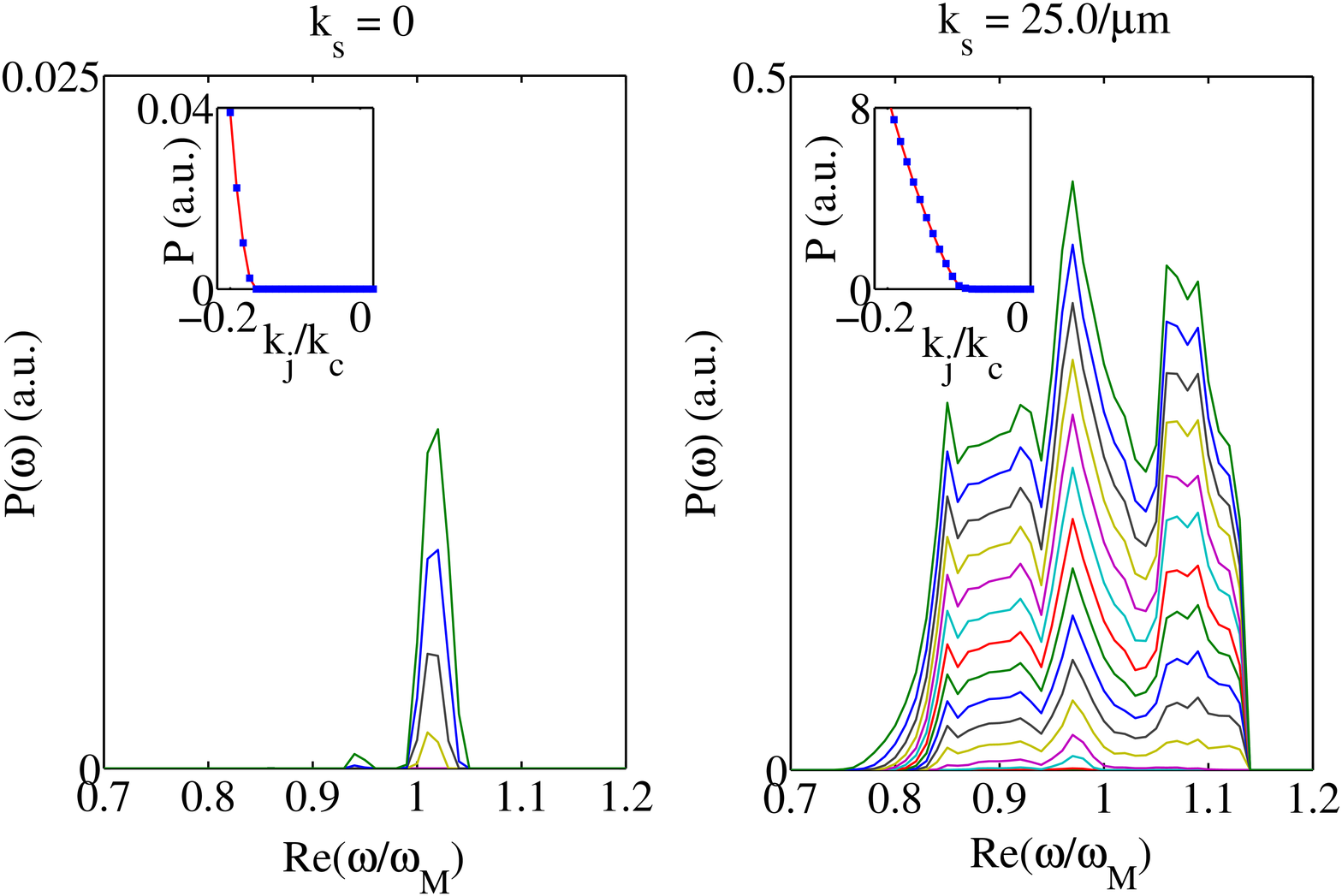}
	\caption{(Color online) Top: Power spectrum (resolution ${\delta}{\omega}/{\omega}_M = 0.01$) at various current
	levels ($k_j = 0.2k_c$ from the top decreasing by ${\Delta}k_j = 0.01k_c$) without (left: $k_s =
	0$) and with (right: $k_s = 25.0$/${\mu}$m) surface anisotropy. Inset: the integrated power
	versus $k_j$.
%	Note $x$-axis is slightly different for the left and right panels, the $y$-scale is about
%	10-100 times smaller than the left panels.
	} 
	\label{fig:Pw}
\end{figure}
%-----------------------------------------

%Power spectrum.
We introduce an approximate power spectrum (\Figure{fig:Pw}) that summarizes all information
about the mode-dependent current-induced amplification:
%-----------------------------------------
\begin{equation}
	P({\omega}) = {\sum}_n{\int}_{\rm Im~{\omega}_{\it n}<0} \abs{\rm Im~{\omega}_{\it n}(\qq)} {\delta}[{\omega}-Re~{\omega}_{\it n}(\qq)]d\qq
\label{eqn:Pw}
\end{equation}
%-----------------------------------------
with $n$ the band index is the density of states at frequency ${\omega}$ weighted by its
amplification. Without surface anisotropy, only a few modes are excited even at a relatively
large current ($k_j = 0.2k_c$). However, when $k_s = 25$/${\mu}$m, the excitation is strongly
enhanced by more than two orders of magnitude due to the easily excitable surface spin wave
modes. Furthermore, we observe broadband excitation over a much larger range of frequencies.
This power spectrum is rather smooth, while the experiments by Kajiwara {\it et al}.
\cite{kajiwara_transmission_2010} show a large number of closely spaced peaks. The latter fine
structure is caused by size quantization of spin waves due to the finite lateral extension of
the sample that has not been taken into account in our theory since it complicates the
calculations without introducing new physics. The envelope of the experimental power spectrum
compares favorably with the present model calculations.

%This approximate power spectrum is continuous because the structure is assumed to be infinite
%in the latteral direction. When the latteral size is finite, the power spectrum becomes
%discretized and shall mimic the experimental power spectrum in Ref.
%\onlinecite{kajiwara_transmission_2010}, and the envelope of the discretized spectrum shall be
%the same as the continuous spectrum in \Figure{fig:Pw}.

The insets in \Figure{fig:Pw} show the integrated power and allow the following conclusions: 1)
the excitation power is enhanced by at least two orders of magnitude by the EASA; 2) the
critical current for magnetization dynamics is $k_j {\sim} 0.08k_c$ for $k_s = 25/{\mu}$m, which
agrees very well with the estimates from \Eq{eqn:kjc}. This critical current is about one order
of magnitude smaller than that for the bulk excitation ($k_c$), and about half of that for MSW
without surface anisotropy ($k_j = 0.16k_c$). For $k_s = 25/{\mu}$m, it corresponds to $J_c =
3{\times}10^{10}$A/m$^2$ for ${\theta}_H = 0.01$ \cite{mosendz_quantifying_2010} and $3.8{\times}10^9$A/m$^2$ for
${\theta}_H = 0.08$ \cite{ando_electric_2008, liu_spin-torque_2011}. These values are calculated for
a film thickness of $d = 0.61~{\mu}$m, but should not change much for $d = 1.3~{\mu}$m corresponding
to the experiment \cite{kajiwara_transmission_2010}, because the excited spin waves are
localized at the interface. Compared to the original estimate $J_c {\sim} 10^{11{\sim}12}$A/m$^2 $,
the critical current for a surface spin wave excitation is much closer to the experimental
value of $J_c {\sim} 10^9$A/m$^2$ \cite{kajiwara_transmission_2010} (although these experiments
report a very inefficient spin wave absorption in contrast to the present model assumption).

According to \Eq{eqn:kjc}, critical current (excitation power) would be further reduced
(increased) by a larger EASA. Ref. \onlinecite{yen_magnetic-surface_1979} reports an
enhancement of the YIG surface anisotropies for capped as compared to free surfaces. A Pt cover
on a YIG surface \cite{kajiwara_transmission_2010} may enhance the surface anisotropy as well.
As seen from \Figure{fig:ks1}, the surface mode (black band) has group velocity ${\partial}{\omega}/{\partial}\qq$
comparable to that of the MSW. The excited surface spin wave therefore propagate and can be
used to transmit spin information over long distance at a much lower energy cost than the bulk
spin waves.

%The reports by Miron {\it et al.} \cite{miron_perpendicular_2011} and Liu {\it et
%al.} \cite{liu_spin-torque_2011} on current-induced magnetization dynamics in layered
%structures are not addressed here since they use ultra-thin metallic instead of thick
%insulating ferromagnetic films.

%Similar structures as in \Figure{fig:NF} is also used experimentally by Miron {\it et. al.}
%\cite{miron_perpendicular_2011} and Liu {\it et. al.} \cite{liu_magnetic_2011} to study the
%current-induced magnetization dynamics. However, their structures uses metal Co instead of
%insulator YIG and is about 1000 times larger in each dimension. Therefore, the physics in their
%experiments would be quite different from this sutdy.

%This study is based on a structure as in \Figure{fig:NF}, which resembles the experimental
%setup in Ref. \onlinecite{kajiwara_transmission_2010} for the purpose of spin information
%transmission. 

%Nevertheless, both observed the magnetization switching by the
%current flow in the NM (Pt) layer, with explanation using a Rashba spin-orbit interaction and
%the spin Hall effect respectively. Due to the much smaller size, the macrospin model used in
%Ref. \onlinecite{liu_spin-torque_2011} might work in their case. But in the macroscopic
%structures, we have to take into account the spin wave excitation beyond the macrospin as
%studied in this Letter.

In conclusion, we predict that an easy-axis surface anisotropy gives rises to a surface spin
wave mode, which reduces the threshold current required to excite the spin waves and
dramatically increases the excitation power. Multiple spin wave modes can be excited
simultaneously at different frequencies and wave-vectors, thereby explaining recent
experiments. Surface spin wave excitations could be useful in low-power future
spintronics-magnonics hybrid circuits.

%=========================================================================
%{\it Acknowledgment:} 
This work was supported by the National Natural Science Foundation of China (Grant No.
11004036), the special funds for the Major State Basic Research Project of China (No.
2011CB925601), the FOM foundation, DFG Priority Program SpinCat, and EG-STREP MACALO. J. X.
acknowledges the hospitality of the G. B. Group at the Kavli Institute of NanoScience in Delft.

%#########################################################################
\bibliographystyle{apsrev}
\bibliography{all}

\end{document}